\documentclass{article}
\usepackage{sandiego06}
\usepackage{graphicx}
\frompage{000} \topage{000}                                              
\def\rightmark{Jet Physics with Identified Particles }
\def\leftmark{R. Bellwied}

\title{Jet Physics with Identified Particles at RHIC and LHC}
\authors{
{Rene Bellwied}\\[2.812mm]
{\normalsize
Department of Physics and Astronomy, Wayne State University \\
Detroit, MI 48201, USA
}}

\abstract{I will present the latest results on particle identified
spectra and two particle correlations at high transverse momentum
measured with the STAR detector at RHIC. I will compare those
measurements with the projected capabilities for the LHC using the
ALICE detector. The effects of string fragmentation and quark
recombination in the intermediate momentum range will be discussed
in the context of hadron formation in proton-proton and Au-Au
collisions, as well as energy loss mechanisms in the dense partonic
phase. The necessity for gamma-jet and heavy quark jet measurements
will be detailed based on the ambiguity of existing particle
identified results.}


\begin{document}

\maketitle
\setcounter{page}{1}

\section{Introduction}
The measurement of particle identified high momentum spectra and
two-particle correlations can be considered the next step in the
characterization of the dense partonic phase created in relativistic
heavy ion collisions at RHIC. Much emphasis is presently given to
the similarities in the light and heavy quark behavior in the
plasma, but these measurements are detailed in other contributions
to this workshop \cite{dasilva,harris}. Here I will mostly focus on
the flavor dependence in light hadron production, i.e. hadrons
consisting of u,d,s quarks. I will address in particular the
question of baryon versus meson production through fragmentation and
recombination as competing mechanisms. In order to determine the
baryon production mechanism in AA collisions we need to understand
the production in pp first. The main question is whether there is a
difference between the light baryon production in the medium and in
the vacuum. In the following section I will briefly review the
latest particle identified pp results from STAR and then move on to
results from AuAu collisions. The ambiguity in the production
mechanism as deduced from single particle properties such as the
suppression at high pt and elliptic flow v2 can be partially
resolved by analyzing two particle correlations. I will show that
early results of intermediate p$_{T}$ associated particle production
are surprisingly featureless regarding flavor dependencies that were
expected from perturbative QCD calculations in vacuum. In contrast,
the single particle spectra suppression at high p$_{T}$ in AA seems
very much affected by flavor suppression in the proton-proton
system, in particular for strange particles. I will try to argue
that this effect should be unique to the strange quark, only if the
strange quark generation is gluon dominated and the heavy quark
production is quark dominated. Many of the studies at RHIC lack the
statistics to map out the transition from medium effects to pure
vacuum fragmentation. I will show that these measurements are
readily accessible at the LHC, in particular with the ALICE
detector.


\section{Why is pp so important ?}

The latest strange and non-strange hadron production results
obtained in proton-proton collisions by STAR, were presented at this
workshop \cite{heinz,ruan}. They show that simple leading order
fragmentation codes are not sufficient to describe the baryon
production at RHIC energies. Next to leading order calculations do
well as soon as quark separated fragmentation functions are used
\cite{akk}, in other words there are contributions in the baryon
spectra from non-valence quark fragmentation. In a leading order
calculation this fact can apparently be approximated by increasing
the K-factor and the multiple parton scattering contribution to the
underlying event \cite{bellwied1}. These effects are expected to be
significantly less pronounced at higher incident energies
\cite{eskola}. In other words, next to leading order contributions
are more important at RHIC energies than at LHC energies.

One important new pp result is the break-down of the so called
m$_{T}$-scaling \cite{gatoff,schaffner} at sufficiently high
transverse mass at RHIC. Fig.1a shows the measured spectra of all
particle identified species obtained by STAR. After appropriate
scaling of each spectrum the m$_{T}$-scaling plot (Fig.1b) can be
devised. Clearly the m$_{T}$-scaling at lower m$_{T}$, which had
previously been established through measurements at lower energies
\cite{isr1,isr2}, can be confirmed. The noticeable deviation from
simple scaling at sufficiently high m$_{T}$ (m$_{T}$ $>$ 3 GeV/c) is
a new feature, though. The spectra seem to group into common baryon
and meson curves at these momenta. This is a surprising effect in
elementary pp collisions which is nevertheless described
quantitatively in PYTHIA, as shown in Fig.2.

\begin{figure}
\begin{minipage}{15pc}
\begin{center}
\includegraphics[width=2.in]{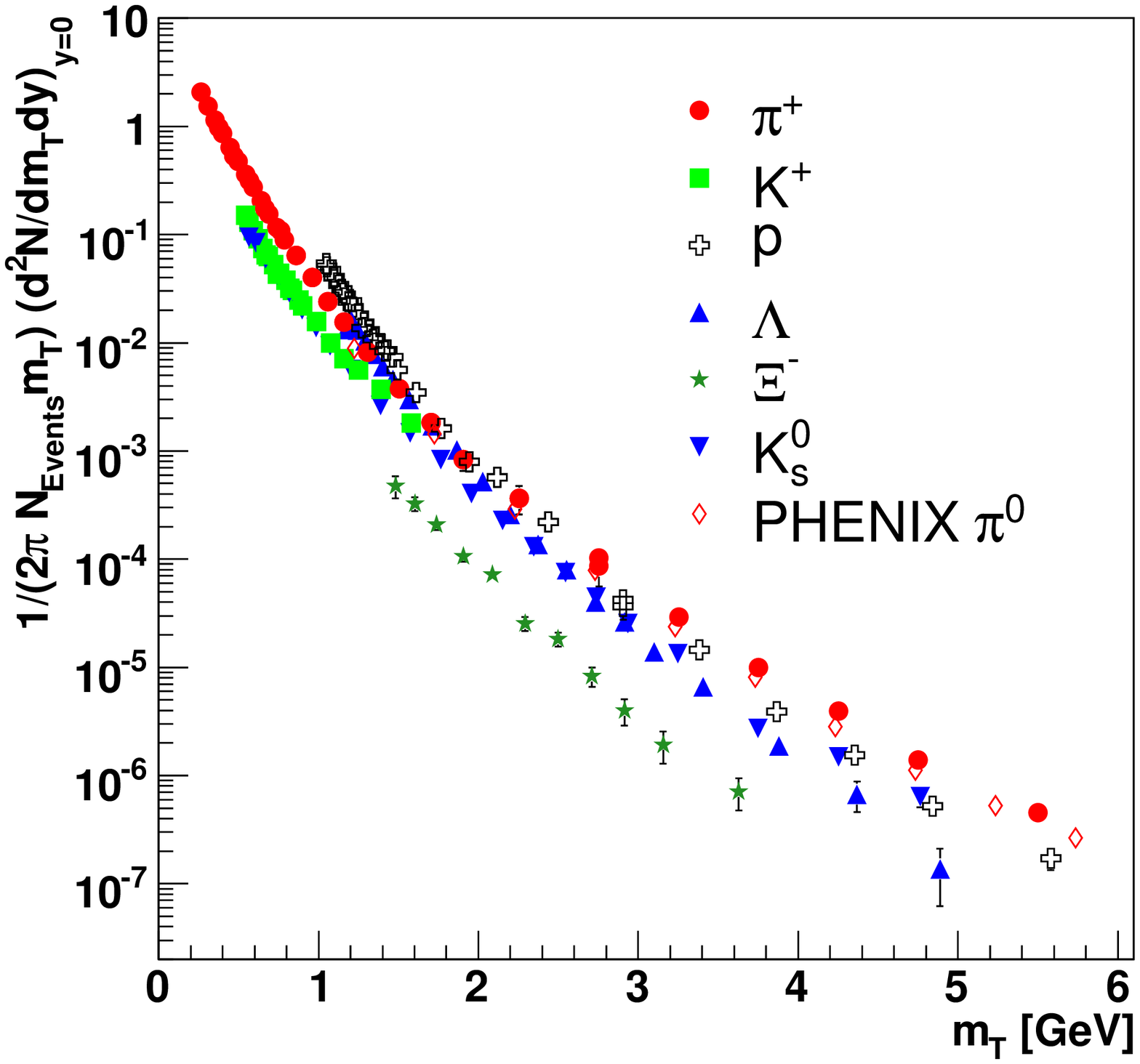} \label{fig:1a}
\end{center}
\end{minipage}\hspace{1pc}%
\begin{minipage}{15pc}
\begin{center}
\includegraphics[width=2.in]{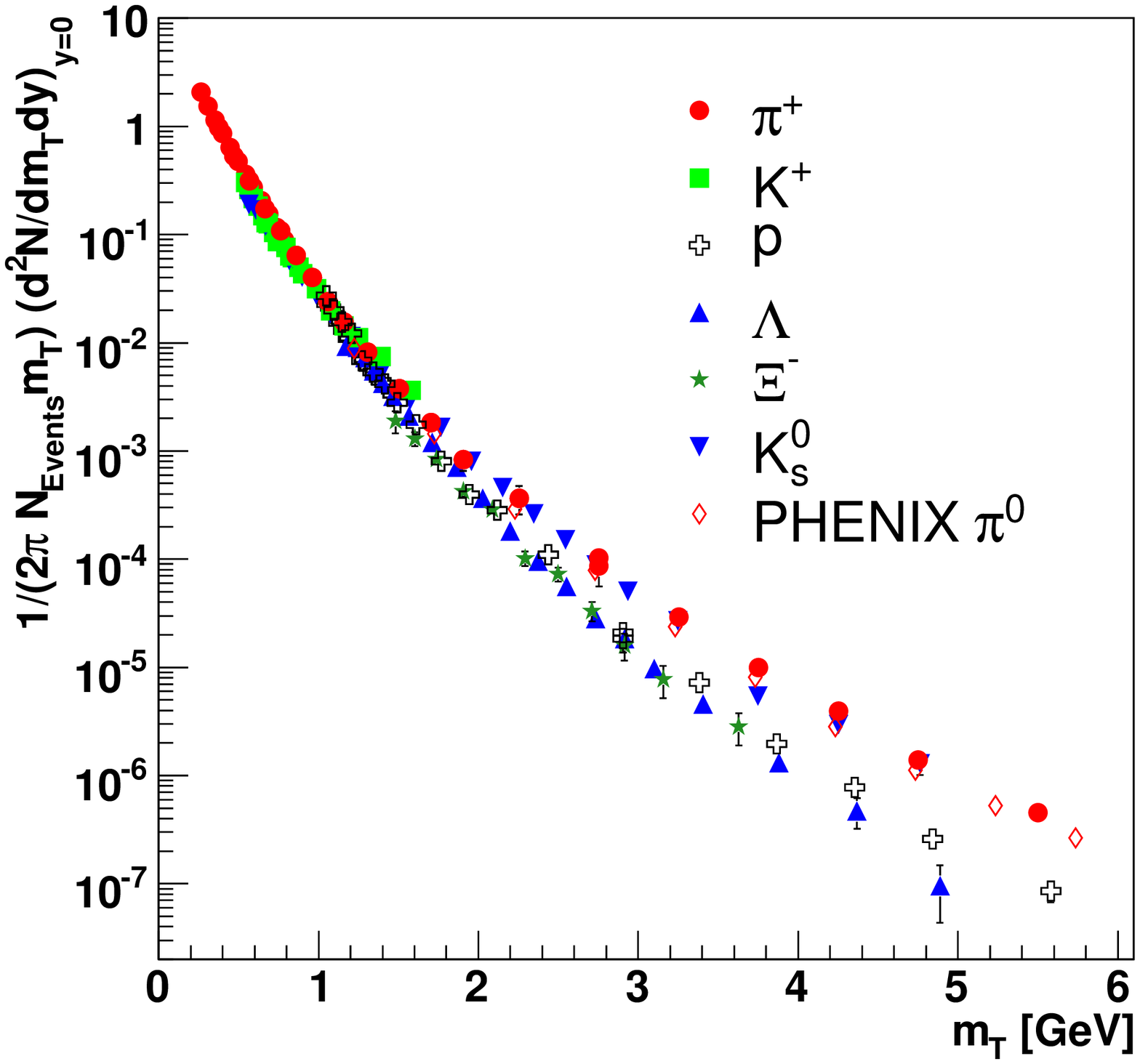}  \label{fig:1b}
\end{center}
\end{minipage}
\caption{(a) Identified transverse mass spectra as measured by STAR
in pp collisions. (b) Same spectra as in Fig.1a but scaled via
multiplicative factors to the measured pion spectrum.}
\end{figure}

Figs.2a and 2b break down the effect, as modeled in PYTHIA, for
gluon and quark jets. The combined spectrum (Fig.2c) is dominated by
the gluon jets. According to PYTHIA the ratio of gluon to quark jets
is about 2:1 at RHIC energies.

\begin{figure}
\begin{minipage}{9pc}
\begin{center}
\includegraphics[width=1.5in]{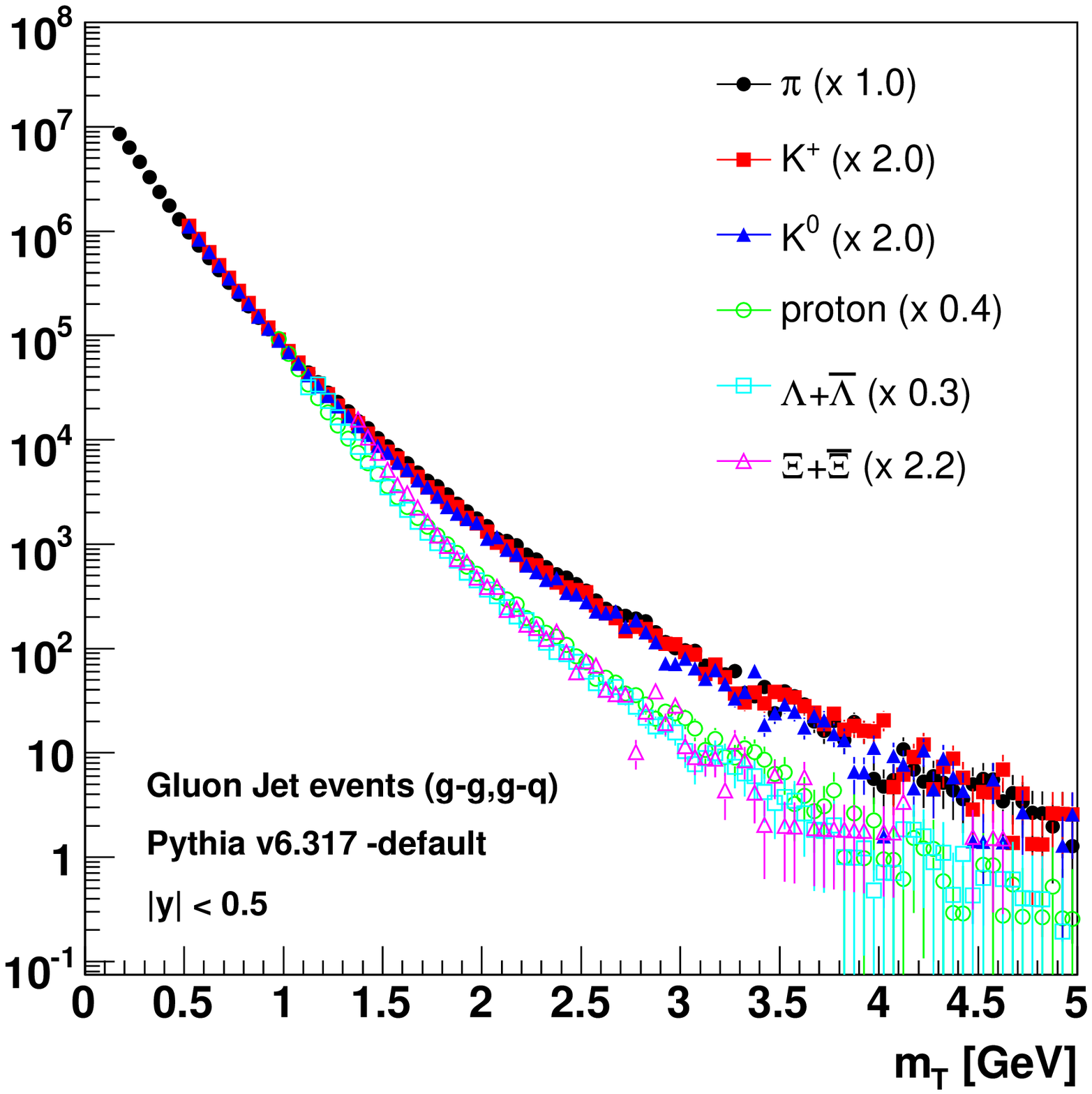} \label{fig:2a}
\end{center}
\end{minipage}\hspace{1pc}%
\begin{minipage}{9pc}
\begin{center}
\includegraphics[width=1.5in]{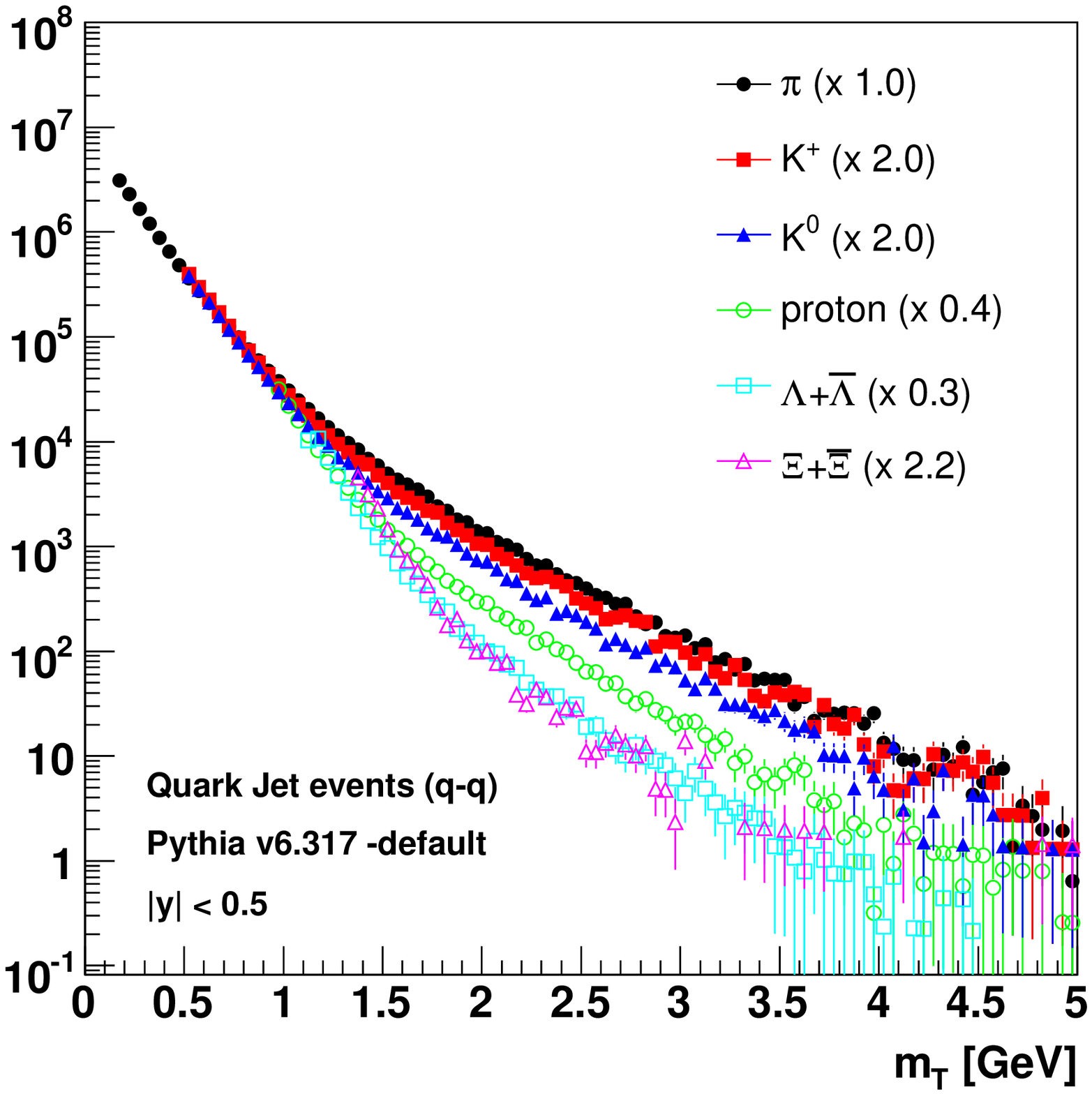} \label{fig:2b}
\end{center}
\end{minipage}\hspace{1pc}%
\begin{minipage}{9pc}
\begin{center}
\includegraphics[width=1.5in]{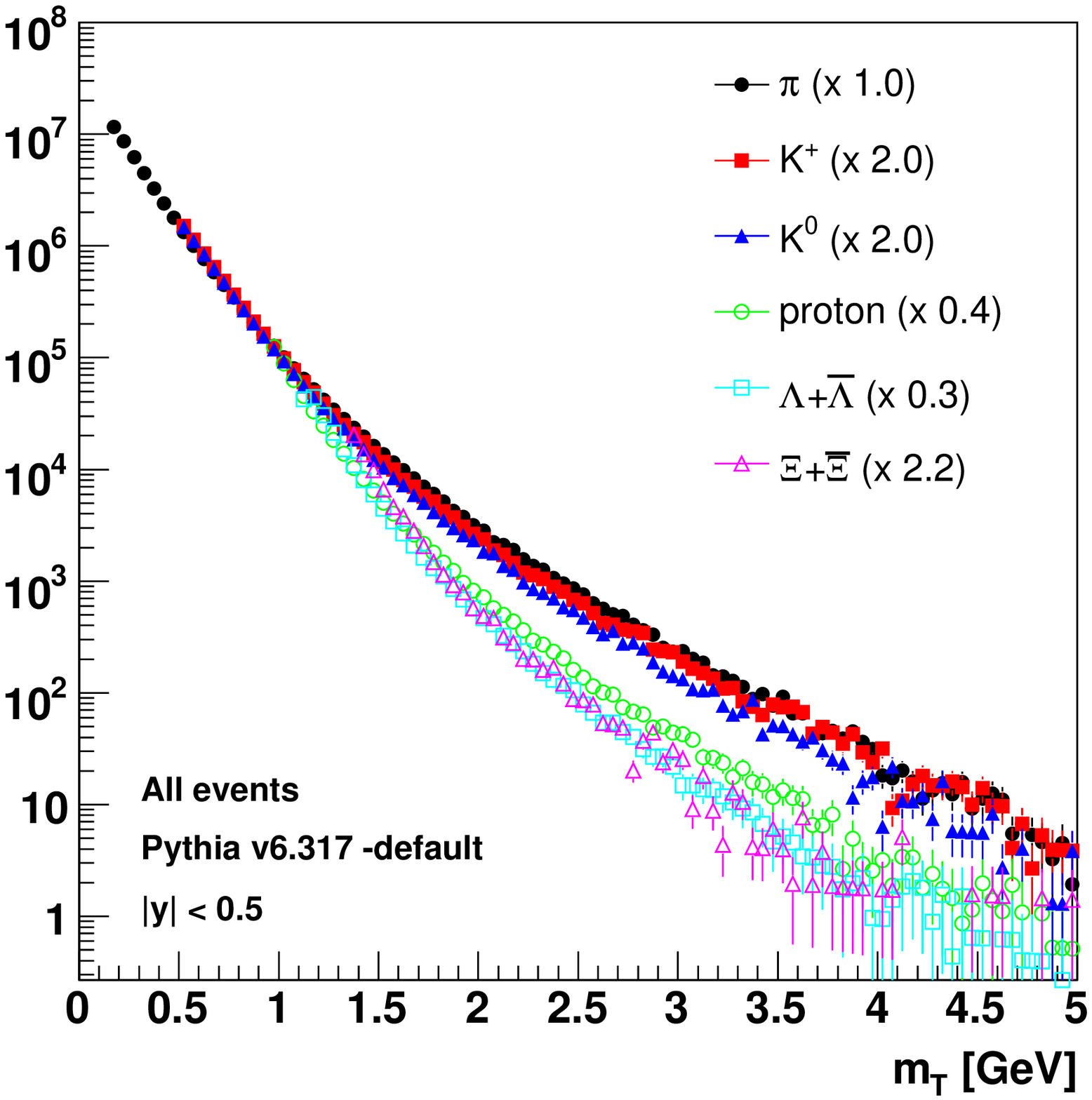} \label{fig:2c}
\end{center}
\end{minipage}
\caption{PYTHIA simulation of the scaled identified mt spectra in
Fig.1b from (a) gluon jet fragmentation (g-g or q-g) only, (b) quark
jet fragmentation (q-q) only, and (c) sum of all parton
fragmentation processes based on the relative ratio of quark to
gluon jets at 200 GeV according to PYTHIA.}
\end{figure}

One very interesting feature in the parton separated spectra
(Figs.2a and 2b) is that apparently the gluon jet fragmentation
leads to a baryon/meson difference at high m$_{T}$, whereas the
quark jets show a mass ordering of the high m$_{T}$ spectra. Gluon
fragmentation dominates at RHIC (and even more at LHC) energies, and
based on PYTHIA the baryon/meson splitting effect in the kinematic
spectra is due to the di-quark formation process which is a
pre-requisite for baryon formation in LUND type string fragmentation
\cite{lund-baryon}. The initial di-quark formation leads to a
lowering of the $<$p$_{T}$$>$ for the baryons (the so called
di-quark suppression factor), whereas for mesons the simple
quark-antiquark hadronization does not affect the $<$pT$>$. This
feature in the fragmentation of elementary gluon jets could be
considered the seed of the baryon meson differences in AA
collisions. The effect by itself is not big enough, though, to
quantitatively describe the strong peak in the baryon/meson ratio at
intermediate pt in AA, but the overall feature (a bump at
intermediate p$_{T}$) can already be seen in pp collisions and might
be simply enhanced by additional effects in AA. In AA one needs the
combined effects of radial flow and quenching to 'pile-up' the
baryons over mesons in the intermediate p$_{T}$ range. Still it is
important to note that the baryon/meson differences already have
their origin in the basic fragmentation process in pp collisions.
This measurement also hints at the validity of the fragmentation
process as modeled by PYTHIA. It is my opinion that in its nature
the diquark-quark formation process is similar to a three quark
coalescence process and thus exhibits equivalent features.

Another conclusion from the pp modeling is that, although early pQCD
calculations predicted a stronger drop in the anti-baryon over
baryon ratio, our results are in accordance with PYTHIA calculations
if one takes into account the strong gluon dominance at RHIC
energies as shown in Fig.3.

\begin{figure}
\begin{center}
\includegraphics[width=2.5in]{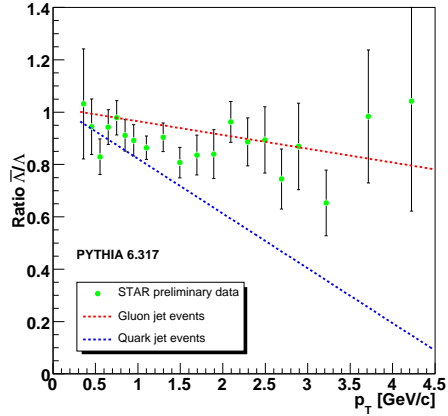} \caption{Measured Anti-$\Lambda$
over $\Lambda$ ratio compared to PYTHIA ratio from gluon and quark
jets.} \label{fig:3}
\end{center}
\end{figure}

In order to unambiguously determine the predicted drop, which is due
to the difference between quark and gluon distribution functions and
the relative contribution of these partons to the baryon and
anti-baryon production, one needs measurements at higher p$_{T}$.
The momentum range of the spectra will be greatly enhanced at the
LHC. Fig.4 shows a projection of p$_{T}$-ranges for identified
spectra based on a scaled LHC pQCD calculation. Generally the light
quark spectra will reach out to at least 20 GeV/c in the first year
of running, which allows for a more unambiguous study of the flavor
effects at high p$_{T}$.

\begin{figure}
\begin{minipage}{15pc}
\begin{center}
\includegraphics[width=2.5in]{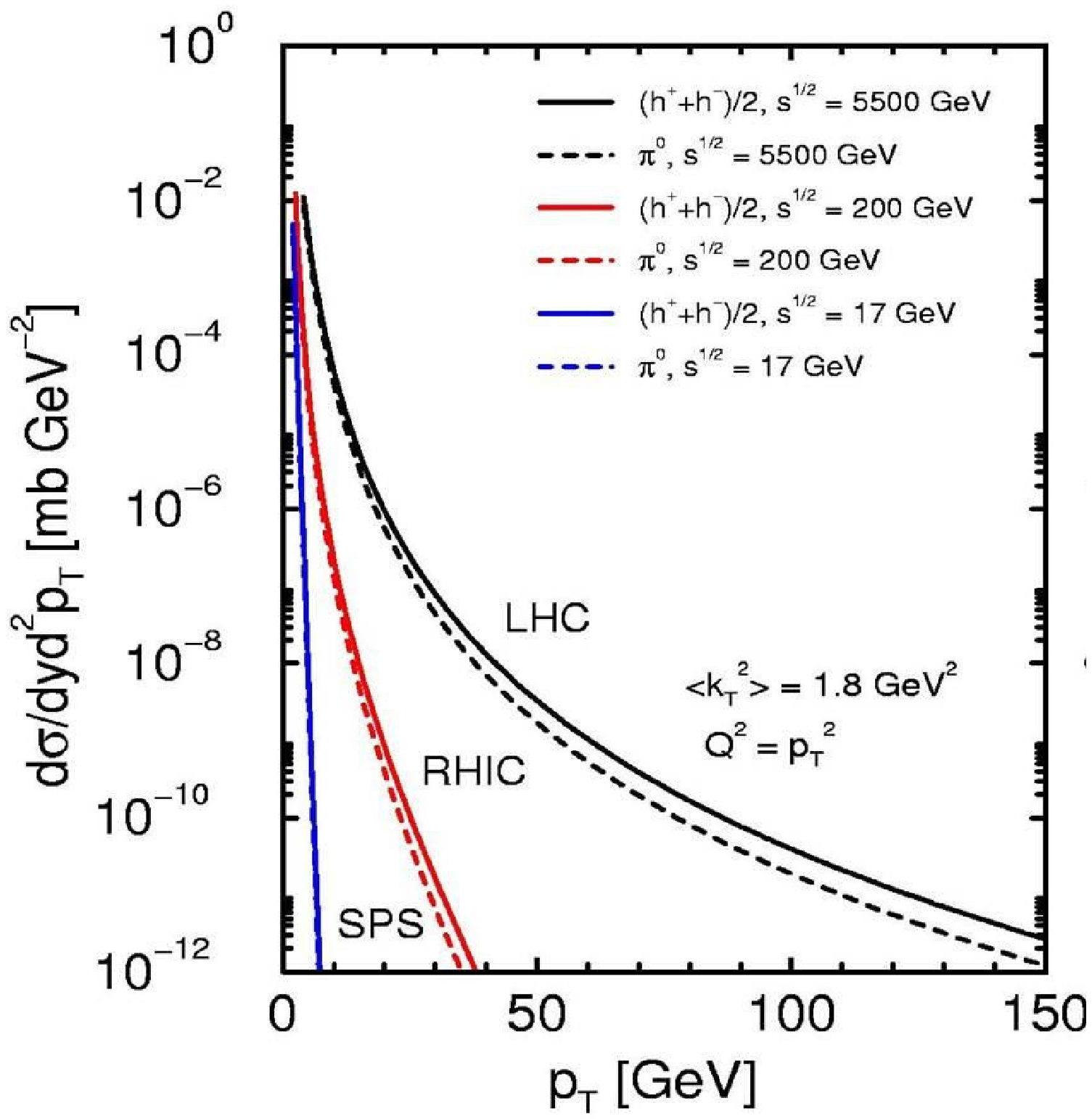} \caption{Leading particle pt-range
based on pQCD calculations for SPS, RHIC, and LHC energies.}
\label{fig:5}
\end{center}
\end{minipage}\hspace{1pc}%
\begin{minipage}{15pc}
\begin{center}
\includegraphics[width=2.5in]{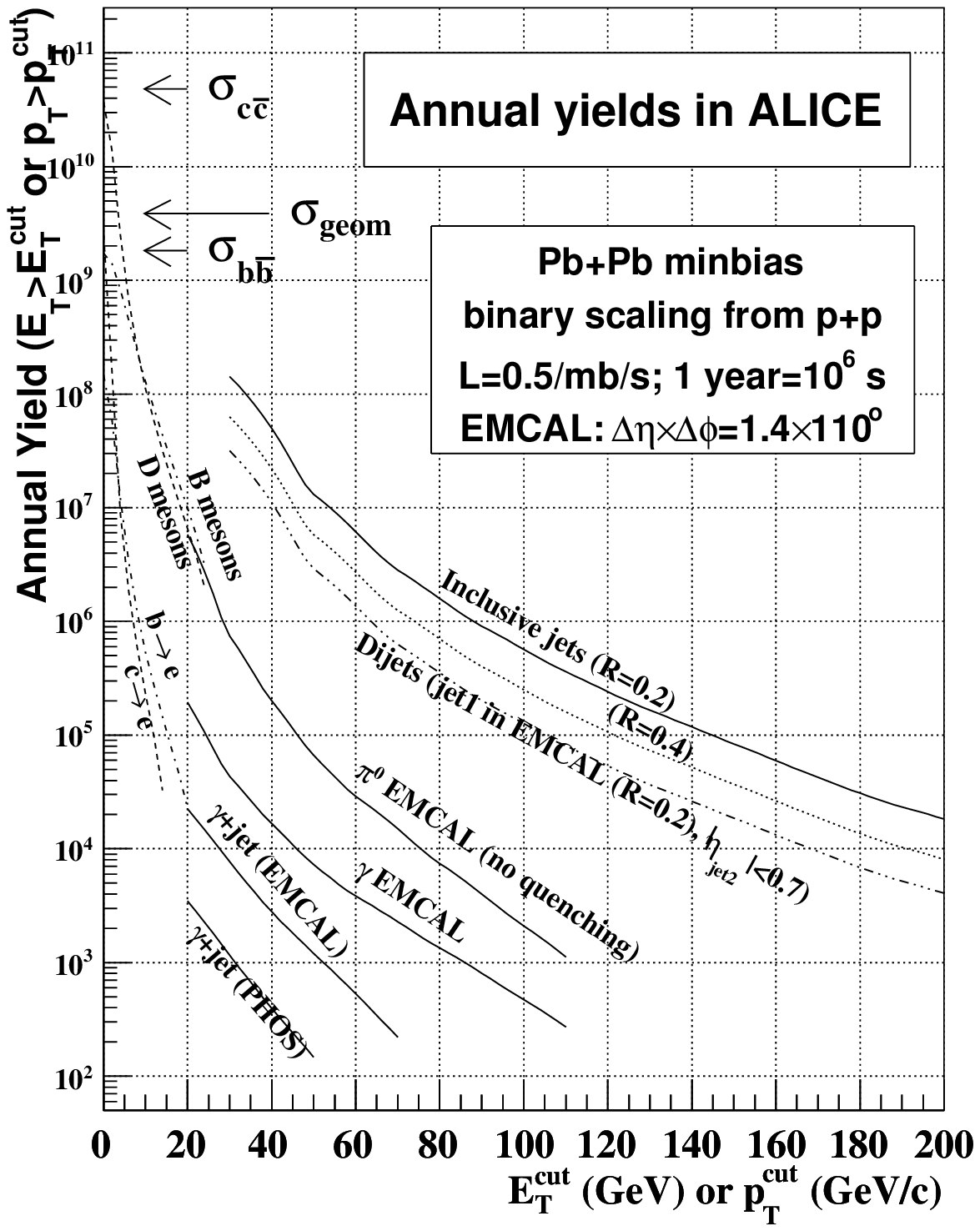} \caption{Annual hard process yields
expected in the ALICE Calorimeter acceptance for minimum bias Pb-Pb
collisions at 5.5 TeV.} \label{fig:6}
\end{center}
\end{minipage}
\end{figure}

One interesting feature in the gluon dominated regime at RHIC and
the LHC is that the $\gamma$-jet process is dominated by Compton
scattering, which leads to a $\gamma$-quark-jet combination in the
outgoing channel. This means not only is the tagged $\gamma$ the
'standard candle' for the jet energy measurements, but it is also a
good trigger for quark-jet events in the gluon dominated regime. The
rate, based on early simulations using the ALICE calorimeter, is
significantly limited compared to di-jet events, but one can still
expect on the order 10,000 events with a $\gamma$ energy above 50
GeV. A similar quark-jet selection can be achieved by triggering on
heavy mesons as leading particles. Projected rates for all jet
measurements with the proposed ALICE EMCal are shown in Fig.5.

\section{Flavor dependencies in identified high pt AA spectra}

One of the most surprising results of the past year was the apparent
difference between R$_{AA}$ and R$_{CP}$ measurements, in particular
for strange baryons. Figs.6 and 7 show a direct comparison of
strange and non-strange baryons and mesons.

\begin{figure}
\begin{minipage}{15pc}
\begin{center}
\includegraphics[width=2.in]{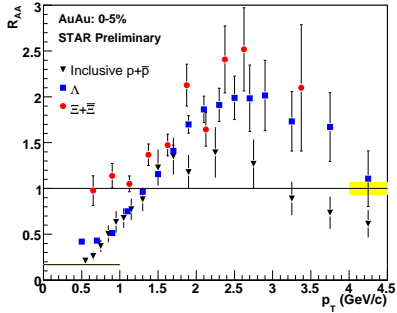} \caption{Nuclear suppression
factor (R$_{AA}$) as measured by STAR.} \label{fig:5}
\end{center}
\end{minipage}\hspace{1pc}%
\begin{minipage}{15pc}
\begin{center}
\includegraphics[width=2.in]{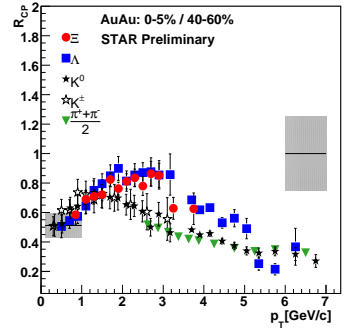} \caption{Nuclear suppression
factor (R$_{CP}$) as measured by STAR.} \label{fig:6}
\end{center}
\end{minipage}
\end{figure}

The strong jet quenching effect at high p$_{T}$ and the slight
baryon/meson splitting due to recombination at moderate p$_{T}$ as
shown on the R$_{CP}$ plot is not visible in the R$_{AA}$ plot. The
R$_{CP}$ suppression has been established for all particle species
from pions to charmed mesons, and besides the already mentioned
baryon/meson difference at intermediate p$_{T}$ the suppression is
surprisingly flavor independent. The R$_{AA}$ pattern though
exhibits a very strong flavor dependence, in particular for baryons.
An enhancement of the high p$_{T}$ yield, rather than a suppression
compared to the pp spectrum, actually increases as a function of the
strangeness content in baryons. A similar enhancement pattern was
previously measured for integrated particle yields, and is generally
attributed to canonical suppression of strange quarks in small
systems \cite{tounsi}, but it is unexpected that apparently this
effect of a small correlation volume in an equilibrated system
should also affect the high p$_{T}$ particle production. Not only
does it lead to an enhancement of the intermediate p$_{T}$ yield
from pp to AA but there is also no evidence of quenching in the
strange baryon R$_{AA}$ plot at high p$_{T}$. This effect is
actually not seen in preliminary results of charmed meson
suppression, so it could indeed be unique to the strange quarks. A
measurement of a charmed baryon ($\Lambda_c$) is needed to
unambiguously determine the difference between the quark flavors.
Because of the difference between R$_{CP}$ and R$_{AA}$ the strange
quark flavor effects need to have their origin in the particle
production in the pp system. This is a good indicator that even
intermediate p$_{T}$ strange baryons are predominantly produced
through coalescence from a thermalized partonic system. In other
words the initial gluon dominated scattering processes leads to
thermalized strange quarks which coalesce into strange baryons. The
effect of the correlation volume during hadronization is still
dominant even at rather large (up to 3 GeV/c) transverse momentum.
Beyond the intermediate p$_{T}$ range the spectrum gets quenched,
but it is still enhanced in AA collisions compared to the pp system.
The question is whether this strange particle production mechanism
drives the pp to AA comparison even in the pure fragmentation regime
above 7 GeV/c. The RHIC experiments do not have a big enough reach
to establish an answer. Fig.4 shows that these measurements can be
achieved at the LHC, though. We know from e$^{+}$e$^{-}$ experiments
that the strangeness suppression factor in the quark condensate is
about 0.4 \cite{e+e-}, but this is just the relative quark
production probability in the hadronization sea, which should also
exist in the medium. In addition, the strangeness saturation factor
increases from pp to AA by a factor two at RHIC energies based on
measurements of integrated strange particle yields \cite{cleymans}.
The factor in AA can be described quantitatively in lattic QCD
\cite{gavai}.

The difference between baryons and mesons in R$_{CP}$ (Fig.7), which
is generally attributed to either recombination or an interplay
between radial flow and jet quenching \cite{lamont}, can also be
described by the so-called Corona effect, which was shown elsewhere
at this conference \cite{werner,pantuev}. The principle here is that
the formed medium consists of a dense core, which follows
hydrodynamics and a corona of pp interactions dominated by multiple
scattering. The main reason for such a distinction and the strong
contribution from the corona is the relative diffuseness of the
nuclear surface, which is not well described by hard spheres. The pp
interactions can be modeled by codes such as EPOS which take into
account the increased parton cascade activity in the low momentum
sector. In these models effects such as baryon/meson splitting in
v$_{2}$ occur because the corona, which carries very little v$_{2}$,
has a much stronger contribution to the light particle spectrum than
the heavy particle spectrum, so it pulls down the hydro v$_{2}$ for
mesons to lower values than for baryons.

Finally one can measure identified two particle correlations at
intermediate p$_{T}$ in order to detect flavor dependencies that are
expected from simple fragmentation arguments. The correlations shown
in Fig.8 show surprisingly little trigger particle flavor
dependence. Again it seems that non-fragmentation processes, such as
recombination, dominate in this p$_{T}$ range. There is evidence for
long-range correlations in $\Delta\eta$ which could be due
production mechanisms that do not exhibit the flavor dependencies of
simple fragmentation. These correlations lead to a significant
enhancement of the associated yield for any trigger species over the
associated yields measured in pp, which are in agreement with the
PYTHIA simulations shown in Fig.8b.

\begin{figure}[hbtl]
\begin{center}
\includegraphics[width=3.in]{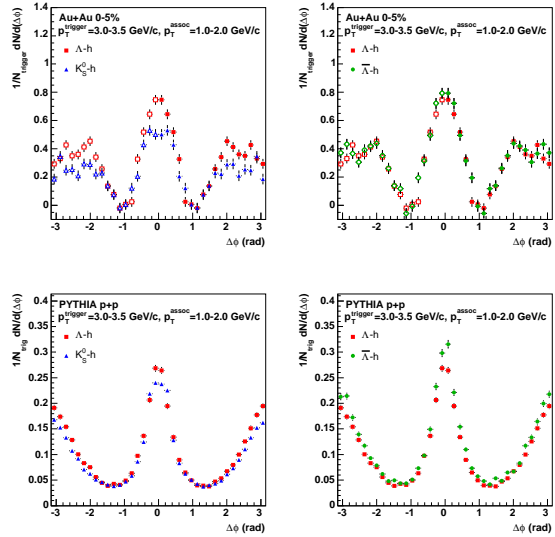} \caption{High p$_T$ two particle correlations
using identified trigger particles and charged hadron as associated
particles. a.) measurements in 0-5\% centrality Au-Au collisions in
STAR, b.) PYTHIA simulations of the same correlations in pp
collisions.} \label{fig:7}
\end{center}
\end{figure}

\section{Summary}

The interpretation of our pp collision results reveal that the
baryon production yields require either multiple scattering through
a soft particle production model such as EPOS or NLO corrections in
pQCD models such as PYTHIA. It is interesting to note that the basic
string fragmentation differences in baryon and meson production lead
to a breakdown of the universal m$_{T}$-scaling of identified
particle spectra. Apparently this breakdown is driven by the baryon
production mechanism in gluon jets and manifests itself as a slope
difference at high m$_{T}$ when comparing baryon to meson spectra.
This basic effect in pp is not sufficient though to describe the
large baryon over meson yield enhancement at intermediate p$_{T}$ in
AA collisions.

Besides the baryon/meson difference there is a surprising absence of
strong flavor effects in the particle to anti-particle ratios in AA,
the identified two particle correlations in AA, and even the jet
quenching in the medium. Thus, it is still an open question whether
the partonic energy loss in AA shows the expected Casimir factor
when comparing hadrons from a fragmenting gluon jet to a quark jet,
i.e. is the energy loss really non-Abelian ?

The only strong flavor effect is in the strangeness sector. High
p$_{T}$ strange baryon production in AA is enhanced instead of
suppressed compared to pp . This could be due to simple canonical
suppression in pp. This thermodynamic effect, which is due to a
limited strangeness phase space occupancy, has been measured for the
first time as a function of transverse momentum, and it is obvious
that the effect is not limited to low momentum or simply bulk
properties. Surprisingly the intermediate p$_{T}$ part scales well
with the canonical suppression factors, which indicates that the
hadronization mechanism of strange baryons, even at higher p$_{T}$,
is driven by a correlation volume, which is distinctly different
from charmed meson production. The D-meson yield and high p$_{T}$
suppression factors in AA are consistent with scaled hard scattering
cross section, i.e. production from string fragmentation.

In identified two particle correlations in AA collisions we see a
strongly enhanced associated particle yield compared to pp,
independent of the trigger particle species. Long range $\Delta\eta$
correlations might account for that and they might be due to
recombination \cite{hwa1}. A small baryon/meson trend can be found
in those correlations but the effect is not very significant. Larger
predicted effects for $\phi$ and $\Omega$ triggered correlations
\cite{hwa2} are under investigation.

In summary, I believe that studies of the hadronization mechanism at
RHIC and LHC energies in vacuum and in medium hold the key to the
puzzle of baryonic matter formation in the universe. We need to
first understand the basic baryon production mechanism(s) in pp
(string fragmentation vs. recombination, di-quark formation ?). Then
we need to determine whether the baryon production mechanism in AA
collisions is modified from the vacuum production. For a more
detailed study the high pt reach and the particle identification
properties of the LHC detectors are crucial.

\section*{Acknowledgements}

I thank Helen Caines, Mark Heinz, and Klaus Werner for useful
discussions.


\vfill\eject

\begin{thebibliography}{99}

\bibitem{dasilva} C.L. DaSilva, contribution to these proceedings

\bibitem{harris} J. Harris, contribution to these proceedings

\bibitem{heinz} M. Heinz, contribution to these proceedings

\bibitem{ruan} L. Ruan, contribution to these proceedings

\bibitem{akk} S. Albino et al., hep-ph/0502188

\bibitem{bellwied1} R. Bellwied for the STAR collaboration, QM05
proceedings, nucl-ex/0511006

\bibitem{eskola} K. Eskola et al., Nucl. Phys. A713 (2003)

\bibitem{gatoff} G. Gatoff and C. Y. Wong, Phys. Rev. D 46, 997 (1992)

\bibitem{schaffner} J. Schaffner-Bielich et al., arXiv:nucl-th/0202054

\bibitem{isr1} P.V. Chliapnikov and V.A. Uvarov, Phys.Lett. B345, 313, 1995

\bibitem{isr2} M. Szczekowski, Phys.Lett. B359, 387, 1995

\bibitem{lund-baryon} P. Eden, G. Gustafson, Z. Phys. C75, 41 (1997)
and hep-ph/9606454

\bibitem{tounsi} K. Redlich and A. Tounsi, Eur. Phys. J. {\bf C24} (2002) 589

\bibitem{e+e-} G. Abbiendi et al. (OPAL), Eur. Phys. J. C16, 407
(2000) and hep-ex/0001054

\bibitem{cleymans} J. Cleymans, J. Phys. G28, 1575 (2002)

\bibitem{gavai} R.V. Gavai and S. Gupta, Phys. REv. D73, 014004
(2006)

\bibitem{lamont} J. Adams et al. (STAR), nucl-ex/0601042

\bibitem{werner} K. Werner, contribution to these proceedings

\bibitem{pantuev} V. Pantuev, contribution to these proceedings

\bibitem{hwa1} C.B. Chiu, R. Hwa, Phys. Rev. C72, 034903 (2005)

\bibitem{hwa2} R. Hwa, C.B. Yang, nucl-th/0602024

\end{thebibliography}
\end{document}